\documentclass[acmsmall,authorversion,nonacm]{acmart}

\usepackage{fancyhdr}
\AtBeginDocument{%
    \fancyfoot[L]{\textit{\textbf{Preprint for CHASE2024.\\  
    https://doi.org/10.1145/3641822.3641868
    }}}%
}

\acmConference[CHASE 2024]{17th International Conference on Cooperative and Human Aspects in Software Engineering}{April 2024}{Lisbon, Portugal}
\usepackage{booktabs}
\usepackage[normalem]{ulem}
\usepackage{tabularx}

\copyrightyear{2024}
\acmYear{2024}
\setcopyright{rightsretained}
\acmConference[CHASE '24]{2024 IEEE/ACM 17th International Conference on Cooperative and Human Aspects of Software Engineering }{April 14--15, 2024}{Lisbon, Portugal}
\acmBooktitle{2024 IEEE/ACM 17th International Conference on Cooperative and Human Aspects of Software Engineering (CHASE '24), April 14--15, 2024, Lisbon, Portugal}\acmDOI{10.1145/3641822.3641868} \acmISBN{979-8-4007-0533-5/24/04}

\acmSubmissionID{13965.6}

\usepackage{xcolor}
\newcommand{\red}[1]{{\textcolor{black}{#1}}}
\newcommand{\blue}[1]{{\textcolor{black}{#1}}}

\begin{document}

\title{Challenges in Understanding the Relationship between Teamwork Quality and Project Success in Large-Scale Agile Projects}

\author{Torgeir Dingsøyr}
\affiliation{
\institution{Norwegian University of Science and Technology}
\city{Trondheim}
\country{Norway}
}

\email{torgeir.dingsoyr@ntnu.no}

\author{Phillip Schneider}
\affiliation{
\institution{Technical University of Munich}
\city{Munich}
\country{Germany}
}
\email{phillip.schneider@tum.de}

\author{Gunnar Rye Bergersen}
\affiliation{
\institution{University of Oslo}
\city{Oslo}
\country{Norway}
}

\email{gunnab@ifi.uio.no}

\author{Yngve Lindsjørn}
\affiliation{
\institution{University of Oslo}
\city{Oslo}
\country{Norway}
}

\email{ynglin@ifi.uio.no}

\renewcommand{\shortauthors}{Dingsøyr, Schneider, Bergersen and Lindsjørn.}
\begin{abstract}
A number of methods for large-scale agile development have recently been suggested. 
Much of the advice in agile methods focuses on teamwork.
Prior research has established that teamwork quality influences project success both for traditional software development teams and agile teams. 
Further, prior studies have also suggested that teamwork quality may play out differently in large projects compared to small. 
We investigated the relationship between teamwork quality and project success with a survey of~196 project participants across~34 teams in four projects, replicating a previous study on single teams.

The new data do not fit the previously established theoretical model, which raises several concerns. 
The \blue{observed}
effect of teamwork quality on project success operates differently across projects.

We discuss possible reasons, which include disagreements on what characterises success in large-scale agile development, "concept drift" of teamwork quality factors, the possibility that interteam factors might have more influence on project success than intrateam factors, and finally, that our study design does not capture all relevant levels and functions.

We conclude with a call for more studies on the quality and frequency of interaction between teams in addition to internal team factors to further advance theory and practice within large-scale agile software development.

\end{abstract}

\maketitle

\section{Introduction} 
\label{section:introduction}
Although most advice from fields such as project management and software engineering is to avoid large software development projects, such projects often are needed and indeed critical for organisations \cite{dingsoyr2019}.
Large-scale software development projects are increasingly using agile methods, and we are starting to see a number of empirical studies focusing on challenges and success factors for such projects \cite{edison2021, uludag2022}. 

Development method is one of many factors which have an influence on project success \cite{mcleod2011}. A key characteristic of agile development methods is that the methods focus on teamwork \cite{baham2021}.
Previous studies have shown that teamwork quality has a significant impact on software project success, both for traditional development \cite{Hoegl2001} and for agile development teams \cite{Lindsjorn2016}, but that teamwork quality affects project success differently in small and large projects \cite{Lindsjorn2018}.

%Building on a previous survey of teamwork quality,
We were interested in exploring teamwork in the context of larger projects, in large-scale agile development, as many organisations put much emphasis on selecting a method \cite{conboy2019}, and we know little of how such methods influence teamwork quality and subsequently project success.
Our research question was, \textit{does the effect of teamwork quality on project success differ between large-scale agile development methods?}
In the following, we provide background and then describe the results from this exploratory analysis. We had access to four projects with data from a total of 34 teams.
The main result was, however, that our data do not fit the previously established theoretical model. 
We discuss concerns with the measurement of teamwork quality and project success.
Finally, we conclude this short paper with recommendations for future studies to explore teamwork in the context of large-scale agile development.

\section{Teamwork quality and multiteam projects} 
\label{section:foundations}
To provide a background, we describe prior work on teamwork quality and project success, teamwork quality in multiteam projects, and finally, large-scale agile development methods and teamwork.

\subsection{Teamwork Quality and Project Success} 
\label{section:teamworkquality_s}
There is a large body of studies on teamwork in general and a growing number of studies of teamwork in software development and in the particular context of agile software development \cite{strode2022}. 
Procedures for teamwork are described in software development methods, for example, on daily meetings in Scrum and programming in pairs in extreme programming. There are also many other factors which will 
affect project success 
for software teams~\cite{mcleod2011}. 
Teamwork is important, however, as it can usually be influenced at project initiation, for example, in staffing teams with a facilitator and requiring teams to conduct regular demonstrations and retrospectives. Such interventions are usually easier than, for example, changing the staffing in a project.

\begin{figure}[t]
  \centering
  \includegraphics[width=\columnwidth]{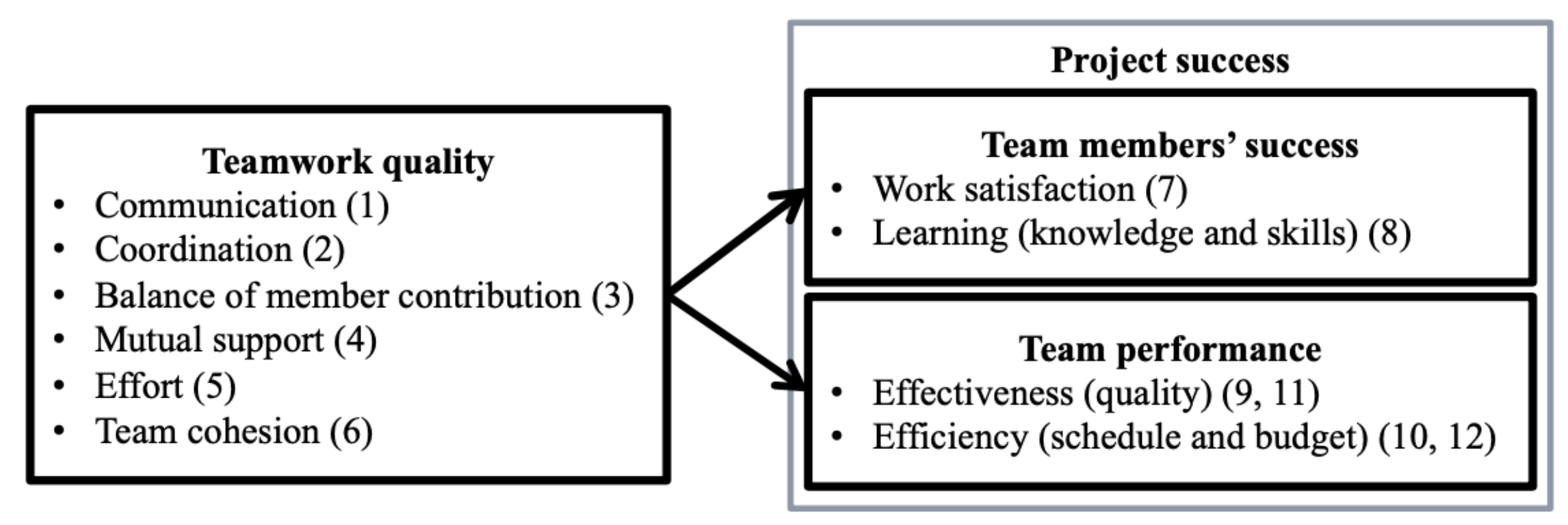}
  \caption{Theoretical model, modified from~\cite{Hoegl2001} and \cite{Lindsjorn2016}, used as conceptual model in analysis}
  \label{fig:GenericProcess}
\end{figure}

One of the general studies of teamwork quality with the most impact has been the survey study on the relationship between teamwork quality and project success in 145 traditional software teams by Hoegl and Gemuenden~\cite{Hoegl2001}. This study established six teamwork quality (TWQ) constructs, which were \textit{communication}, \textit{coordination}, \textit{balance of member contributions}, \textit{mutual support}, \textit{effort}, and \textit{cohesion}. Project success was measured as \textit{personal success} and \textit{team performance}.
A modified theoretical model is shown in Figure~\ref{fig:GenericProcess} where \textit{teamwork quality} affects \textit{team member´s success} and \textit{team performance}.

A replication study by Lindsjørn et al.~\cite{Lindsjorn2016} with 71 teams found that the relationship also held in agile software development when team members and team leaders rated team performance. 
A negligible effect was found for product owners. Comparing the effect of teamwork quality on project success, it was found to be only marginally greater for agile teams than traditional ones in the original survey, despite the focus on teamwork in agile software development. 

A recent mixed method study with quantitative data from two thousand scrum teams by Verwijs and Russo~\cite{verwijs2023}, finds that scrum team effectiveness depends on five high-level factors: responsiveness, stakeholder concern, continuous improvement, team autonomy, and management support. 

\begin{table*} [btbp]
    \centering
  \caption{Characteristics of data collection from four projects (CD=Continuous development)}
  \label{tab:dataset-overview}
  \begin{tabular}{lllll}
    \toprule
    \textbf{Characteristic} & \textbf{LeSS1} & \textbf{LeSS2} & \textbf{SAFe1} & \textbf{SAFe2} \\
    \midrule
    Industry & public sector & defense & finance & utilities \\
    Organisation size & small & large & small & large \\
    Team background & national & national & national & international \\
    Program increment &   11 of 12 &  3 of 4 &  3 in CD & 13 in CD \\
    Agile method & Scrum/XP & Scrum/XP & SAFe 4.6 essential & SAFe 4.6 portfolio \\
    Work style & co-located & partly remote & partly remote & fully remote \\
    Country & Norway & Norway & Germany & Germany \\
    \midrule
    No. of teams & 11 & 10 & 8 & 5 \\
    No. of respondents & 78 & 42 & 55 & 21 \\
    %\red{\sout{Mean number of respondents per team}} & \red{\sout{7.1}} & \red{\sout{4.2}} & \red{\sout{6.9}} & \red{\sout{4.2}} \\
    \bottomrule
  \end{tabular}
\end{table*}

\subsection{Teamwork Quality in Multiteam Projects} 
\label{section:teamworkquality_b}
Projects with multiple teams have been studied in software development and other domains. A study from the automotive industry found that collaborative processes such as interteam coordination, project commitment, and teamwork quality have "predictive properties in regard to later team performance and can serve as early warning indicators"~\cite{Hoegl2004}. % p. 38
The study also identified a positive relationship between interteam coordination, project commitment, and teamwork quality.

In the context of large-scale agile development, we find several studies investigating teamwork quality, for example, in modifying the teamwork quality instrument to the context of agile software development \cite{poth2020, poth2021}.

A multiteam project can be seen as a portfolio of related projects. A study on how portfolio management practices, such as focusing on a business case existence, strategic clarity, and operational control, was found in one study to constrain the relationship between teamwork quality and project success \cite{bechtel2021}.

Other teamwork-related factors such as psychological safety, team cohesion, knowledge sharing, team reflexivity, and team structure are proposed as influencing team performance and team learning \cite{ahmad2022}.

\subsection{Large-scale Agile Development Methods and Teamwork} 
\label{section:teamworkquality}
Large-scale agile development methods provide advice on how to organise product development with many development teams \cite{dingsoyr2019, edison2021}. Workshops with researchers have put emphasis on topics such as interteam coordination, interteam communication, and knowledge sharing \cite{bass2019}. Empirical studies such as \cite{dingsoyr2018, dingsoyr2023} and overview studies \cite{edison2021} show a range of extra arenas, work practices, and roles on top of ones that typically exist in agile development teams. This includes arenas for interteam coordination and communication such as scrum of scrums or "product owner coordination meetings", practices such as "big room planning", and roles such as "release train engineer" and "area product owner". 
Batra~\cite{batra2020} discusses quantitative studies of large-scale development and states that "Given that coordination and teamwork are critical success factors, do we need to create new constructs such as synchronising ability?"

\section{Research Method}
\label{section:researchmethod}
We conducted an exploratory quantitative multi-case study, collecting data with a survey as a differentiated replication \cite{Lindsay1993} in order to investigate if the findings from previous studies generalise to teamwork in large software projects. 
While Hoegl and Gemuenden \cite{Hoegl2001} studied traditional software teams, and Lindsjorn et al. \cite{Lindsjorn2016} agile teams, we collected data on agile teams in four large-scale agile software projects. Two Norwegian projects using a method close to Large-Scale Scrum (LeSS) and two German Scaled Agile Framework (SAFe) projects. \blue{LeSS is a "minimal extension" of Scrum for large-scale product development, while SAFe is 
\red{a}
comprehensive method \cite{dingsoyr2019}}. 
Our research approach, from the data collection to the statistical analysis, is detailed in the following subsections 
\blue{and follows the approach in~\cite{Lindsjorn2016}}: 

\subsection{Data Collection and Study Sample}
\label{section:datacollection}
To measure teamwork quality variables in the Norwegian projects, we used the questionnaire reported in Lindsjørn et al.~\cite{Lindsjorn2016}. 
\red{A German translation of the English questionnaire was used for the two German projects.}
Depending on the team distribution in each project, the survey was either conducted at the workplace or as an online survey.
The two collected survey datasets from the Norwegian projects were also used in two previous publications \cite{Lindsjorn2016,Lindsjorn2018}. The German datasets were taken from two master theses at the Technical University of Munich \cite{doepp2019, styrsky2021}.
We merged the datasets, filtering for a team-based inclusion criterion.
From all questionnaire participants, we only included teams that came from a large-scale agile project and for which at least one developer, as well as one product owner, responded. 

As can be seen from Table~\ref{tab:dataset-overview}, 
\blue{our} % to distinguish from Linsjørn et al 2016
study sample consists of 196 respondents from 34 teams working in four projects. 
Because of the team-based inclusion criterion, there are more LeSS teams~(21) than SAFe teams~(13) in the merged dataset. 
The same holds for the number of respondents. 
The number of teams per project ranges from 5 to 11, with a
mean of 5.8 
\blue{respondents} 
\red{per team}.
In general, the team sizes between the projects were comparable. 

In terms of qualitative differences, the four projects differ in various characteristics. For instance, all projects came from different industry sectors: finance, utilities, defense, and the public sector. The projects were executed by two small and two large organisations. The Norwegian projects have finished, while the German are continuous development projects. Most of the projects had teams that were situated in the same country, whereas one project had an international team. Depending on the specific project, the teamwork style was co-located, partly remote, or fully remote. The only similarity between the projects of the same countries was the type of employed agile development method. While the Norwegian teams used a combination of Scrum and Extreme Programming close to LeSS, the German teams applied SAFe.

The data and 
\red{the detailed analyses are available in a replication package.\footnote{\url{https://doi.org/10.5281/zenodo.10571869}}}

\subsection{\blue{Statistical Analysis}}
\label{section:dataanalysis}

\red{We use confirmatory factor and correlation analyses to explore the interactions between the variables under investigation, as previously detailed in~\cite{Lindsjorn2016}.
Specifically, we calculate the arithmetic mean of the responses to each question at the team level (our unit of analysis). 
These means are then aggregated into the variables of interest.}
\blue{There were negligible missing data overall (1.5\%), but nearly all missing data were from the product owners (5.1\%).}
\red{For the analysis, we use the lavaan package~\cite{rosseel2012lavaan} in R for confirmatory Structural Equation Modeling.} 

To test the model fit for the specified measurement and structural model, we report the Root Mean Square Error of Approximation (RMSEA) and its 95\% confidence interval. 
A close model fit is indicated by RMSEA values below 0.05 and values above 0.10 indicate an unacceptable model fit. 
As all indices of model fit have limitations, we also calculated the Comparative Fit Index, the Tucker-Lewis Index, and the Standardised Root Mean Square Residual; all these indices support the inferences we make throughout the sections, but we do not report these. We do, however, report the chi-square statistic ${\chi}^2$, which roughly should be less than two times the degrees of freedom.

\section{Results}
\label{section:results}

\begin{table*}[bt] \centering 
  \caption{Descriptives and correlations for the investigated variables} 
  \label{tab:descr_and_correlations} 
\resizebox{\textwidth}{!}{\begin{tabular}{@{\extracolsep{4pt}} lcc|cccccccccccc} 
\\[-1.8ex]
\hline \\[-1.8ex] 
 & $\overline{X}$ & SD & (1) & (2) & (3) & (4) & (5) & (6) & (7) & (8) & (9) & (10) & (11) & (12) \\ 
\hline \\[-1.8ex] 
Communication$^{\ddagger}$ (1) & $3.99$ & $0.40$ & $1$ & $0.70$ & $0.69$ & $0.69$ & $0.69$ & $0.25$ & $0.69$ & $0.64$ & $0.31$ & $0.31$ & $0.82$ & $0.26$ \\ 
Coordination (2) & $3.87$ & $0.48$ & $0.65$ & $1$ & $0.70$ & $0.83$ & $0.69$ & $0.65$ & $0.65$ & $0.64$ & $0.23$ & $0.24$ & $0.77$ & $0.58$ \\ 
Bal. member contr. (3) & $4.02$ & $0.44$ & $0.70$ & $0.75$ & $1$ & $0.73$ & $0.59$ & $0.65$ & $0.65$ & $0.77$ & $0.64$ & $0.77$ & $0.77$ & $0.49$ \\ 
Mutual support$^{\ddagger}$ (4) & $4.26$ & $0.44$ & $0.70$ & $0.80$ & $0.63$ & $1$ & $0.33$ & $0.65$ & $0.55$ & $0.68$ & $0.77$ & $0.68$ & $0.66$ & $0.32$ \\ 
Effort (5) & $3.94$ & $0.47$ & $0.73$ & $0.72$ & $0.75$ & $0.54$ & $1$ & $0.78$ & $0.31$ & $0.64$ & $0.68$ & $0.64$ & $0.37$ & $0.24$ \\ 
Cohension (6) & $3.94$ & $0.41$ & $0.84$ & $0.75$ & $0.76$ & $0.75$ & $0.76$ & $1$ & $0.24$ & $0.64$ & $0.64$ & $0.64$ & $0.28$ & $0.69$ \\ 
% \hline \\[-1.8ex] 
Work satisfaction$^{\ddagger}$ (7) & $4.21$ & $0.52$ & $0.75$ & $0.67$ & $0.59$ & $0.79$ & $0.65$ & $0.83$ & $1$ & $0.54$ & $0.65$ & $0.55$ & $0.91$ & $0.58$ \\ 
Learning$^{\ddagger}$ (8) & $4.16$ & $0.47$ & $0.62$ & $0.48$ & $0.48$ & $0.66$ & $0.56$ & $0.73$ & $0.91$ & $1$ & $0.55$ & $0.31$ & $0.62$ & $0.45$ \\ 
Effectiveness TM (9) & $3.99$ & $0.42$ & $0.61$ & $0.63$ & $0.65$ & $0.58$ & $0.68$ & $0.81$ & $0.68$ & $0.60$ & $1$ & $0.23$ & $0.53$ & $0.50$ \\ 
Efficiency TM (10) & $3.67$ & $0.52$ & $0.58$ & $0.53$ & $0.58$ & $0.43$ & $0.60$ & $0.67$ & $0.41$ & $0.23$ & $0.71$ & $1$ & $0.34$ & $0.38$ \\ 
Effectiveness PO (11) & $4.05$ & $0.60$ & $0.16$ & $0.32$ & $0.38$ & $0.13$ & $0.46$ & $0.37$ & $0.31$ & $0.25$ & $0.58$ & $0.46$ & $1$ & $0.67$ \\ 
Efficiency PO$^{\ddagger}$ (12) & $3.86$ & $0.84$ & $0.31$ & $0.21$ & $0.32$ & $0.11$ & $0.31$ & $0.46$ & $0.40$ & $0.38$ & $0.51$ & $0.48$ & $0.69$ & $1$ \\ 
\hline \\[-1.8ex] 
\end{tabular}}
\raggedright
\footnotesize{Notes. $\overline{X}$ is the mean and SD the standard deviation. TM is team member and PO is product owner. N = 34 for all variables. The lower diagonal is the observed Pearson correlation and the upper diagonal is the transformed (estimated) correlation matrix using SEM. Correlations above 0.34 are significant at $p < 0.05$, and correlations above 0.44 are significant $p < 0.01$.
Variables that are not normally distributed according to the Shapiro-Wilk test of normality are marked
\blue{$^{\ddagger}$.
Variables 1--6 are treated as independent and variables 7--12 dependent in the analysis.
}}
\end{table*}

Table~\ref{tab:descr_and_correlations} shows the variable descriptives and 
\red{Pearson} 
correlations for the merged dataset. 
Five of the twelve investigated (aggregated) variables have non-normal distributions.
\blue{By visual inspection, the lack of normality for these variables appears to arise from}
outliers. 
\blue{Moreover,}
the aggregated variables associated with teamwork quality (1--6), personal success (7--8) appear (also by visual inspection) to have approximate linear relations, which is an assumption in the statistical analyses we report. 
However, the aggregated variables associated with team performance as rated by the 
the product owners (11--12) 
do not. 

Overall, when inspecting the differences in the patterns of correlations, the product owners using SAFe 
\red{appear to be}
in more agreement (positive correlations) with the team members than in the LeSS projects.

Table~\ref{tab:sem-results-tem} reports model fit for the two different large-scale agile development methods and the merged dataset. The overall lack of fit of the data to the theoretical model is clear from the unacceptable RMSEA values and their confidence intervals. 
In particular, the SAFe data do not fit the overall (measurement and structural) model. 
An approach to improve model fit is to relax model assumptions by, for example, introducing additional parameters in the model, allowing for correlated residual (error) variance, or by estimating only parts of the model, such as the measurement model only. 
Although we could improve the model fit slightly with such approaches, the overall model fit was still clearly unacceptable and consistently worse for the SAFe data. For example, by estimating the measurement model of teamwork quality only for the SAFe data, model fit improved (RMSEA = 0.25 [0.11--0.39], ${\chi}^2$ (9) = 21.24), but it was still unacceptable.

\begin{table}[b]
  \centering
  \caption{Model fit for the theoretical model in Figure~\ref{fig:GenericProcess}}
  \label{tab:sem-results-tem}
  \begin{tabular}{lllr}
    \toprule
    \textbf{Dataset} & \textbf{n} & \textbf{RMSEA [95\% CI]} & \textbf{${\chi}^2$ (df)} \\ 
%    \textbf{Dataset} & \textbf{No. of. Teams} & \textbf{RMSEA} & \textbf{Chi-square}  \\
    \midrule
    LeSS & 21 & 0.15 [0.07--0.22]& 74.86 (51) \\ % NOR
    SAFe  & 13 & 0.42 [0.34--0.49] & 156.43 (51) \\ % GER
    \midrule
    Merged & 34 & 0.17 [0.13--0.22] & 102.19 (51)\\
    \bottomrule
  \end{tabular}
\end{table}

\section{Discussion}
\label{section:discussion}
This study sought to investigate how and if the effect of teamwork quality on team performance differs between large-scale agile development methods. We analysed data from four projects with~5 to~11 development teams, two Norwegian and two German.

The original study by Hoegl and Gemuenden~\cite{Hoegl2001} on the relationship between teamwork quality and project success for innovative projects found a significant positive relationship between teamwork quality and team performance as rated by team members, team leaders, and managers.

In the context of agile software development, a replication of the study by Lindsjørn et al.~\cite{Lindsjorn2016} found a positive effect of teamwork quality on team performance as rated by team members and team leaders.  A negligible effect was found for product owners. Comparing the effect of teamwork quality on project success, it was found to be only marginally greater for agile teams, despite the focus on teamwork in agile software development. 

Further, when comparing the influence of teamwork quality on project success for small and large agile projects, Lindsjørn et al.~\cite{Lindsjorn2018} found that teamwork quality primarily had an impact on product quality in small projects, while teamwork quality had a positive impact on product quality when rated by team members but a negative impact when rated by team leaders. The study further suggests that teamwork quality affects project success differently in small and large projects.

When analysing data from large-scale agile development projects, what could explain why our data do not fit the previously established theoretical model?
We discuss some possible explanations in the following, which include aspects of research design as well as measurements used.

\subsection{Project Success}
There could be larger disagreements as to what characterises success in large-scale than in small-scale software development. It might be difficult to build a shared understanding of the quality of the product and if delivery is according to schedule and budget when the project involves a number of teams who work on their own parts of the product. Although software projects seldom are "megaprojects" (cost of USD one billion or more), large software projects might suffer some of the same challenges as megaprojects as they are risky due to long planning horizons and complex interfaces, decision-making involves multiple stakeholders, and the scope or ambition level can change over time \cite{Flyvbjerg2014}.

A strength of the study design of \cite{Hoegl2001} compared to, for example,~\cite{verwijs2023}, is that the dependent variable is measured by someone outside the team (in our case a product owner).
Prior studies have, however, found that team members rate performance differently than managers, and this is explained by team members giving high-performance ratings if teams have healthy internal processes, while managers who have less insight into team dynamics focus on external factors such as communication with external agents \cite{cohen1997}.
Still, one would then expect that the SAFe data would fit only the measurement model of teamwork quality, as this is only evaluated by the team members; however, these data do not fit the teamwork quality model in isolation either (RMSEA~=~0.25~[0.11--0.39], ${\chi}^2{(9)} = 21.24$).
Following recommendations by \cite{kahneman2021}, it could make sense to ask respondents to rank teams rather than indicating project success on scales. In particular, in a multiteam context, it will often be one product owner who is responsible for many teams, and this person should then have a good background for ranking.

\subsection{Measuring Teamwork Quality}
We have measured teamwork quality with an instrument 
constructed over~20 years ago. 
Clearly, software development practices and processes have changed during these years to mitigate known problems with teamwork and to help deliver better software. 
The recent study by \cite{verwijs2023} 
offers an alternative theoretical model,   
indicating that there might be factors other than the teamwork quality model that better explain project success. The team effectiveness model by Strode et al. \cite{strode2022} is another source of factors.

Thus, one could argue that the understanding of the "teamwork quality" construct may have changed over time. Such a shift over time of what is measured is sometimes referred to as "concept drift"~\cite{gama2014}. From a construct validity perspective~\cite{sjoberg2022}, such a change would involve the threat of ``construct underrepresentation'' because key elements of the teamwork quality that affected team performance 
% are no longer captured in 
is not captured by
the instrument.

One way forward is to use qualitative studies involving teams in an attempt to establish what they (and their product owners) regard as the main drivers of team performance in the context of large-scale agile development.

\subsection{Multiteam Characteristics}
Studies of "teams of teams", "multiteam systems" in organisational psychology, suggest that cross-team (interteam) processes are more important than within-team (intrateam) processes for the overall performance of the "multiteam system" \cite{marks2005}. Unlike~\cite{Hoegl2004}, our survey instrument did not include cross-team factors, which could have a larger impact on project success than what we measure with the teamwork quality instrument at team level. Such factors have been identified in overview studies on large-scale agile development, what Edison et al.~\cite{edison2021} label "connecting practices". Examples of such practices are arenas for interteam coordination, release planning and architecture, and knowledge sharing and improvement. There are, of course, also other factors that will influence project success, such as the skills of developers in important fields (technology and the domain knowledge of the solution) and how the project is able to engage users \cite{mcleod2011}. Also, there may be conflicts between teams or incentive structures that reduce the motivation for interteam collaboration. The importance of additional processes for large-scale development is supported by Rolland et al. \cite{rolland2023}, who argue that agile practices cannot be scaled in a 
"straightforward linear fashion" such as adding a layer with "Scrum of Scrums" on top of daily meetings for coordination and communication, but that additional roles and practices are needed to manage scale.

\subsection{Multiple Levels}
Finally, Batra~\cite{batra2020} discusses challenges and opportunities in quantitative studies of large-scale agile development. He identifies challenges such as operationalising dependent variables and that researchers "may get overwhelmed" when selecting independent variables as there is a "plethora of suggested constructs". A suggestion is to study the phenomenon at different levels of analysis and three functional perspectives. The levels of analysis are organisational, team, as well as individual, and the three functional perspectives are process, governance, and outcome. Our instrument focused on team level and outcome, combined with a description of process and governance.

We could also add time as a factor, as for example, project success might be measured differently when measured during the project or after project completion.

In line with Batra~\cite{batra2020}, we suggest that future quantitative studies of large-scale agile development efforts include more analytical levels and functions. We believe teamwork is a key analytical level, but emphasis should be made on the quality and frequency of the interaction between teams in addition to the internal team factors. Literature reviews on large-scale agile software development ~\cite{edison2021, uludag2022}  show suggestions for independent variables that can be measured.

\subsection{Limitations}
The primary limitation of our analysis lies in the limited number of observations. While we have nearly 200 respondents, the data are grouped into teams, which are then combined into four large-scale projects representing two agile methods. This gives a larger uncertainty to conclusions. 
However, we do claim that there are concerning issues present, an argument we believe can be made with less data than if we were to claim that "the data supports the theoretical model."

Another constraint in our study arises from the diverse nature of the collected data, which varies in at least seven characteristics
(see Table~\ref{tab:dataset-overview}), which all might influence the needs for coordination and communication across teams. This study cannot pinpoint which of the context factors contributes most significantly to project success.
Furthermore, there are likely other crucial contextual factors, such as team capability factors (see, e.g.,~\cite{boehm1981}), that are not represented in our dataset.

\section{Conclusion and Future Work}
\label{section:conclusion}
One of the main topics in software process research in recent years has been "large-scale agile development"~\cite{dingsoyr2019, uludag2022}. 
Many methods are available, and many organisations strive to find a method that fits~\cite{conboy2019}. 
Teamwork is a key characteristic of agile software development. We used a team level study design to focus on multi-team projects using two different methods for large-scale agile development. We were interested in exploring if the use of the different methods would impact the relationship between teamwork quality and project success, which has been established in prior research both for traditional teams~\cite{Hoegl2001} and for agile development teams~\cite{Lindsjorn2016}. 
However, our data did not fit the previously established theoretical model, which raised several concerns. We discussed four possible reasons: Larger disagreements as to what characterises success in large-scale agile development, the teamwork quality instrument might not capture all important aspects of teamwork, interteam factors might have a larger impact on project success than the teamwork quality, and finally, that our model fails to capture important analytical levels and functions.

In line with Batra~\cite{batra2020}, we suggest that future quantitative studies should provide a deeper understanding of the dynamics of large-scale agile development by exploring more analytical levels and functions.
In particular, we recommend studies to focus on the quality and frequency of interaction between teams in addition to internal team factors.

\subsection*{Acknowledgement}
\label{section:acknowledgement}
We would like to thank the anonymous reviewers for their inspiring comments and Dr. Ömer Uludağ, Manuel Styrsky, and Maximilian Doepp for their work in collecting the German data used in this study.

\bibliographystyle{ieeetr}

\bibliography{references}
\end{document}